\documentclass[prl,showpacs,twocolumn]{revtex4}
\usepackage{amsmath,amsfonts,amsthm}
\ifx\pdftexversion\undefined
  \usepackage[dvips]{graphics}
\else
  \usepackage[pdftex]{graphics}
\fi
\newtheorem{thm}{Theorem}
\newcommand{\ket}[1]{\vert{#1}\rangle}

\def\Cx{\mathbb C}

\def\Hil{{\mathcal H}}

\newcommand{\egpic}{
\begin{picture}(230,70)(0,30)
\put(10,80){\line(1,0){20}}  \put(10,80){\line(0,-1){10}}
\put(30,70){\line(-1,0){20}} \put(30,70){\line(0,1){10}}
\put(15,90){\vector(0,-1){10}}
\put(20,80){\vector(0,1){10}}
\put(25,80){\vector(0,1){10}}
\put(40,80){\line(1,0){15}}  \put(40,80){\line(0,-1){10}}
\put(55,70){\line(-1,0){15}} \put(55,70){\line(0,1){10}}
\put(45,90){\vector(0,-1){10}}
\put(50,80){\vector(0,1){10}}
\put(65,80){\line(1,0){25}} \put(65,80){\line(0,-1){10}}
\put(90,70){\line(-1,0){25}} \put(90,70){\line(0,1){10}}
\put(70,90){\vector(0,-1){10}}
\put(75,90){\vector(0,-1){10}}
\put(80,90){\vector(0,-1){10}}
\put(85,80){\vector(0,1){10}}
\put(100,80){\vector(1,0){10}}
\put(120,80){\line(1,0){20}} \put(120,80){\line(0,-1){10}}
\put(140,70){\line(-1,0){20}} \put(140,70){\line(0,1){10}}
\put(125,90){\vector(0,-1){10}}
\put(130,80){\vector(0,1){10}}
\put(145,80){\oval(20,20)[t]}
\put(150,80){\line(1,0){15}} \put(150,80){\line(0,-1){10}}
\put(165,70){\line(-1,0){15}} \put(165,70){\line(0,1){10}}
\put(160,80){\vector(0,1){10}}
\put(175,80){\line(1,0){25}} \put(175,80){\line(0,-1){10}}
\put(200,70){\line(-1,0){25}} \put(200,70){\line(0,1){10}}
\put(180,90){\vector(0,-1){10}}
\put(185,90){\vector(0,-1){10}}
\put(190,90){\vector(0,-1){10}}
\put(195,80){\vector(0,1){10}}
\put(210,80){\vector(1,0){10}}
\put(20,40){\line(1,0){20}} \put(20,40){\line(0,-1){10}}
\put(40,30){\line(-1,0){20}} \put(40,30){\line(0,1){10}}
\put(25,50){\vector(0,-1){10}}
\put(30,40){\vector(0,1){10}}
\put(45,40){\oval(20,20)[t]}
\put(50,40){\line(1,0){15}} \put(50,40){\line(0,-1){10}}
\put(65,30){\line(-1,0){15}} \put(65,30){\line(0,1){10}}
\put(70,40){\oval(20,20)[t]}
\put(75,40){\line(1,0){25}} \put(75,40){\line(0,-1){10}}
\put(100,30){\line(-1,0){25}} \put(100,30){\line(0,1){10}}
\put(85,50){\vector(0,-1){10}}
\put(90,50){\vector(0,-1){10}}
\put(95,40){\vector(0,1){10}}
\put(110,40){\vector(1,0){10}}
\put(130,40){\line(1,0){20}} \put(130,40){\line(0,-1){10}}
\put(150,30){\line(-1,0){20}} \put(150,30){\line(0,1){10}}
\put(135,50){\vector(0,-1){10}}
\put(167.5,40){\oval(55,35)[t]}
\put(155,40){\oval(20,20)[t]}
\put(160,40){\line(1,0){15}} \put(160,40){\line(0,-1){10}}
\put(175,30){\line(-1,0){15}} \put(175,30){\line(0,1){10}}
\put(180,40){\oval(20,20)[t]}
\put(185,40){\line(1,0){25}} \put(185,40){\line(0,-1){10}}
\put(210,30){\line(-1,0){25}} \put(210,30){\line(0,1){10}}
\put(200,50){\vector(0,-1){10}}
\put(205,40){\vector(0,1){10}}
\end{picture}}

\begin{document}
\title{A Ferromagnetic Lieb-Mattis Theorem}
\author{Bruno Nachtergaele\footnote{
Supported by the National Science Foundation (DMS-0303316).}}
\affiliation{Department of Mathematics, University of California at Davis, Davis, CA 95616}
\email{bxn@math.ucdavis.edu}
\author{Shannon Starr\footnote{Supported in part by NSERC.}}
\affiliation{Department of Mathematics, McGill University, Quebec, Canada H3A 2K6}
\email{sstarr@math.mcgill.ca}
\date{\today}
\begin{abstract}
We prove ferromagnetic ordering of energy levels for XXX Heisenberg
chains with spins of arbitrary magnitude, thus extending our previous result
for  the spin 1/2 chain. Ferromagnetic ordering means that the minimum energies in the invariant subspaces  of fixed total spin are monotone decreasing as a function of the total spin.
This result provides a ferromagnetic analogue of the well-known theorem by
Lieb and Mattis about ordering of energy levels in antiferromagnetic and ferrimagnetic systems
on bipartite graphs.
\end{abstract}
\maketitle
\section{Introduction and main result}\label{sec:introduction}\label{sec:main}

\vspace{-2.5pt}
A famous theorem of Lieb and Mattis \cite{LM},
with a subsequent remark by Lieb in \cite[footnote 6]{Lieb},
proves ``ordering of energy levels'' for a large class of
Heisenberg models on bipartite lattices.
Namely, if the two sublattices are $A$ and $B$, and all
interactions within $A$ and $B$ are ferromagnetic
while interactions in between $A$ and $B$ are antiferromagnetic,
then the unique ground state multiplet has
total spin equal to $|\mathcal{S}_A - \mathcal{S}_B|$,
where $\mathcal{S}_A$ and $\mathcal{S}_B$ are the
maximum total spins on the two sublattices.
Moreover, the minimum energy in the invariant subspace
of total spin $S$, for $S\geq |\mathcal{S}_A - \mathcal{S}_B|$,
is monotone increasing as a function of $S$.
An important example is the usual antiferromagnet
on a bipartite lattice with equal-size sublattices.
Then the ground state is a unique spin singlet,
and the minimum energy levels for each possible total spin $S$, are
monotone increasing in $S$.

In this paper we derive a theorem about the ordering of energy levels for a
class of {\em ferromagnetic} Heisenberg models. For a ferromagnet
we expect the minimum energy for fixed total spin $S$ to be
monotone decreasing in $S$. In particular, this
agrees with the well-known fact that the ground state has maximum total
spin, and also with the usually tacit assumption
that the first excited state must have total spin equal to the maximal
value minus one. On the regular lattices this
can be deduced by an exact spin-wave calculation.
On arbitrary graphs, or with spins of varying magnitudes,
exact results are not available but our argument  proves that
monotone ordering holds at least in one-dimensional models.

Our main object of study is the ferromagnetic Heisenberg chain
of $L$ spins ($L\geq 2$). The spins may vary from
site to site: $s_x \in \{\frac{1}{2},1,\frac{3}{2},2,\ldots\}$,
for $x=1,\ldots,L$. Moreover, we allow arbitrary, positive coupling constants
$J_{x,x+1}>0$. The quantum Hamiltonian is thus,
\begin{equation}
H = - \sum_{x=1}^{L-1} J_{x,x+1} \left[\frac{1}{s_x s_{x+1}}
\boldsymbol{S}_{x} \cdot \boldsymbol{S}_{x+1}-1\right]\,.
\label{eq:ham}\end{equation}
Here, $S^i_{x}$ ($i=1,2,3$) denote the standard spin-$s_x$ matrices at site $x$,
and $\boldsymbol{S}_{x} \cdot \boldsymbol{S}_{x+1} = \sum_{i=1,2,3} S_{x}^i S_{x+1}^i$.
Without loss of generality, we normalize the spin matrices by the magnitudes, $s_x$, for later convenience, and subtract a constant so that the ground state energy vanishes.

Since the Hamiltonian is $\textrm{SU}(2)$ invariant, the vectors of a given
total spin span an invariant subspace for $H$. We call $S$
an admissible spin for $H$ if this subspace contains
at least one nonzero vector. For admissible spins, we define $E(H, S)$ to
be the smallest eigenvalue of $H$ among all those with eigenvectors of total spin $S$.
(We may define it to be infinite if $S$ is not admissible.)
Clearly, the largest admissible spin for $H$ is $\mathcal{S}:=\sum_{x=1}^L s_x$.

We say that (strict) ferromagnetic ordering of energy levels (FOEL) holds if
it is true that $E(H,S) \geq E(H,S')$ (with strict inequality) whenever $S<S'$
are a pair of admissible spins for $H$.
\begin{thm}\label{thm:main}
Strict ferromagnetic ordering of energy levels holds for every
nearest-neighbor spin chain, with Hamiltonian
as in (\ref{eq:ham}).
\end{thm}
For the special case of the spin-$\frac{1}{2}$ chain this result was proved in \cite{NSS}. We conjecture that all ferromagnetic XXX models, i.e., on {\em any} finite graph, have the FOEL property. In
particular we expect FOEL for the model on a lattice in any dimension \footnote{Even though we cannot
treat the higher-dimensional case at the moment, one should not give up hope
too quickly; it took more than four decades to generalize the Lieb-Schultz-Mattis Theorem \cite{LSM}
about the gap in isotropic Heisenberg models with a unique ground state
(e.g, antiferromagnets) to higher dimensions \cite{Hastings}.}.
In the case of constant spins, $s_x\equiv s$, and constant couplings,
$J_{x,x+1}\equiv J>0$, on a lattice, a spin-wave calculation
implies that the lowest excitations have total spin equal
to $\mathcal{S}-1$. Therefore, since the ground states have maximal total spin,
one has
$$
E(H,S) \geq E(H,\mathcal{S}-1) > E(H,\mathcal{S})\, ,
$$
for all $S< \mathcal{S}$. We are not aware of any previous
conjectures and certainly no proofs for other values of $s$ or for
the  general case with arbitrary $s_x$ and $J_{x,x+1}>0$.

As a major step in the proof of Theorem 1, we will prove another monotonicity
property of the energies $E(H,S)$. Namely, suppose that $H^\prime$ is the Hamiltonian of
the form (\ref{eq:ham}) for a ferromagnetic spin chain obtained by adding one or more spins to
one of the ends of a spin chain with Hamiltonian $H$, and let $\mathcal{S}^\prime$ and $\mathcal{S}$ denote the maximum total spin of these spin chains respectively. Then, for any $0\leq n\leq
\mathcal{S}$, we have
\begin{equation}
E(H,\mathcal{S}-n)>E(H^\prime,\mathcal{S}^\prime-n)
\label{mono1}\end{equation}
Note that here the deviation, $n$, of the maximum total spin is kept fixed.
The inequality (\ref{mono1}), which is part of Theorem \ref{thm:mono} below, has an appealing interpretation if we assume
that the eigenstates corresponding to the minimum energy of fixed total spin
describe a droplet of opposite magnetization (for the spin $1/2$ XXZ chain with
either open or periodic b.c., this has been proved in \cite{NSa,Ken}).
(\ref{mono1}) then states that the ground state energy of such a
droplet decreases with increasing volume, much as if
the droplet behaves as a free particle.

\section{The Proof}

\vspace{-2.5pt}
We will prove Theorem \ref{thm:main} by an induction argument
in which the system is built up in increments of total-spin $\frac{1}{2}$.
The total spin can be increased by $\frac{1}{2}$ in two ways:
either a new spin $\frac{1}{2}$ can be appended to one end of the chain;
or else the magnitude of the spin at the end of the chain can be increased by $\frac{1}{2}$.
We call these case (I) and case (II), respectively.
In this way, we obtain a sequence of models
$H_1,H_2,\dots,H_N$, for $N = 2 \mathcal{S}$,
and with $L_k$ spins of magnitude $s_{x,k}$, which are of the form
\begin{equation}
H_k = - \sum_{x=1}^{L_k-1} J_{x,x+1} \left[
\frac{\boldsymbol{S}_{x} \cdot \boldsymbol{S}_{x+1}}
{s_{x,k} s_{x+1,k}}-1\right]\, ,
\label{eq:hamnew}\end{equation}
and such that $H_1=0$ for a single spin $\frac{1}{2}$, and
$H_N=H$ is the model of (\ref{eq:ham}).

For any $k$, if $H_{k+1}$ is obtained from $H_k$ as in case (I),
then $\Hil_{k+1} \cong \Hil_k \otimes \Cx^2$.
In case (II), there is an orthogonal projection
$P_k : \Hil_k \otimes \Cx^2 \to \Hil_{k+1}$,
and one may define $\Hil'_{k+1} \cong \Hil_k \otimes \Cx^2$
obtained from $\Hil_k$, as in case (I).
Then $P_k$ projects onto the subspace of $\Hil'_k$
spanned by those vectors which have total spin equal
to $s_{L_k}+\frac{1}{2}$ on the last two sites. The central part in the proof
of Theorem 1 is provided by the following theorem, which includes the monotonicity property
mentioned in  (\ref{mono1}). Here and later on, ${\bf I}$ will denote the
indentity operator on $\Cx^2$.
\begin{thm}\label{thm:mono}
Let $H_1,\dots,H_N$ be a sequence of Hamiltonians
as described above.
For each $k=1,\dots,N-1$, depending on whether $H_{k+1}$ is obtained
from $H_k$ as in case (I) or (II),
one of the following two relations holds
\begin{equation}
H_{k+1}\geq H_k\otimes {\bf I}\quad \textrm{or}\quad H_{k+1} = P_k (H_k\otimes {\bf I} )P_k^*\, .
\label{opineq}
\end{equation}
Moreover, if $S$ is an admissible spin for $H_k$,
then $S+\frac{1}{2}$ is admissible for $H_{k+1}$,
and
\begin{equation}
E(H_{k+1}, S+1/2)<E(H_k,S)\, .
\label{mono}
\end{equation}
\end{thm}
We will prove Theorem \ref{thm:mono} later. Assuming it, we can complete the
proof of Theorem \ref{thm:main}, which we will do next.

{\em Proof of Theorem \ref{thm:main}.}
The proof is by induction.
Of course, $H_1$ satisfies FOEL, simply because there is only one possible spin, $\frac{1}{2}$.
Suppose that $H_k$ satisfies strict FOEL.
We have to show that $H_{k+1}$ also satisfies FOEL.
We will consider cases (I) and (II), separately.

For case (I), suppose that $S<S'$ are such that $S+\frac{1}{2}$
and $S'+\frac{1}{2}$ are admissible for $H_{k+1}$.
Note that since $S'>S$, this means that $S'$ is admissible
for $H_k$. Suppose that $\psi \in \Hil_{k+1}$ is the eigenvector
of $H_{k+1}$ with total spin $S+\frac{1}{2}$,
and energy $E(H_{k+1},S+\frac{1}{2})$.
We may assume that $\psi$ is a ``highest-weight'' vector,
meaning that $S^+ \psi = 0$.
Then, by the Clebsch-Gordan series for $\textrm{SU}(2)$
\cite{Edmonds},
there are two highest-weight vectors
$\psi_1,\psi_2 \in \mathcal{H}_k$, (not both equal to zero)
with total-spins equal to $S$ and $S+1$, respectively, and such that
$\psi=(\psi_1 - {\textstyle \frac{1}{2S+2}} S^- \psi_2)\otimes\ket{\uparrow} +  \psi_2\otimes\ket{\downarrow}$.
Using equation (\ref{opineq}), we deduce the energy inequality
\begin{equation*}
\begin{split}
&\langle \psi,H_{k+1} \psi \rangle\,
    \geq\, \langle \psi, (H_k\otimes {\bf I}) \psi \rangle\\
& =\, \langle{\psi_1,H_k \psi_1}\rangle
    + (1 + (2S+2)^{-1})\langle{\psi_2,H_k \psi_2}\rangle\, .
\end{split}
\end{equation*}
Cross terms of the form  $\langle \psi_1, H_k S^- \psi_2\rangle$ vanish since $\psi_1$
and $\psi_2$ are vectors of fixed but different total spin and $H_k$ commutes with the total spin.
The RHS is a convex combination of the energies of $\psi_1$ and $\psi_2$ with respect to $H_k$.
These energies are no less than the minimum of $E(H_k,S+1)$
and $E(H_k,S)$.
But, since $H_k$ satisfies FOEL, this minimum is never less than
$E(H_k,S')$ because $S < S+1\leq S'$.
Hence,
$E(H_k,S)\leq E(H_{k+1},S+\frac{1}{2})$.
However, then using (\ref{mono})
we deduce that
$E(H_{k+1},S'+\frac{1}{2})<E(H_{k+1},S+\frac{1}{2})$
which establishes the induction step.

In case (II), we use the previous argument and
the Rayleigh-Ritz variational principle.
Namely, from (\ref{opineq}),
we conclude that $H_{k+1} = P_k H_{k+1}' P_k^*$,
where $H_{k+1}' = H_k \otimes {\bf I}$ is a
Hamiltonian on $\Hil'_{k+1}$.
By the previous argument,
we conclude that $E(H_{k+1}',S+\frac{1}{2}) \geq E(H_k,S')$.
But by the variational principle, $E(S+\frac{1}{2},H_{k+1}) \geq E(S+\frac{1}{2},H_{k+1}')$
since the Hamiltonians are the same, but in
$E(S+\frac{1}{2},H_{k+1})$ one minimizes over a smaller subspace:
the range of $P_{k+1}$.
Therefore, $E(H_{k+1},S +\frac{1}{2}) \geq E(H_k,S')$, and
the rest of the argument follows case (I). This concludes the proof of
Theorem \ref{thm:main}.

{\em Proof of Theorem \ref{thm:mono}.}
(\ref{opineq}) can be seen as follows.
In case (I), where $H_{k+1}$ is obtained by appending a spin-$\frac{1}{2}$,
it is clear that $H_{k+1} - H_k\otimes {\bf I} \geq 0$ since the difference
is a Heisenberg interaction involving the new spin,
which is non-negative by construction.
In case (II), a simple calculation shows
$H_{k+1} = P_k^* (H_k\otimes {\bf I} )P_k$.
Our choice of spin normalization in (\ref{eq:ham}) is
precisely so as to make this relation hold without additional
numerical factors.

To prove (\ref{mono}), we first observe that $E(H_k,S)$ can be calculated in the sector of
{\em highest weight} vectors of total spin $S$.
These are vectors $\psi \in \Hil_k$, of total spin $S$, such that $S^+ \psi=0$,
or equivalently $S^3_{\textrm{tot}} \psi = S \psi$.
We will calculate the matrix of $H_k$ restricted
to this subspace in a special basis.
In this basis, we find that all off-diagonal
matrix elements are non-positive.
Let $d(k,S)$ denote
the dimension of the highest weight space of spin $S$ for the system
$H_k$, which is also equal to the multiplicity of the spin-$S$ representation
in the chain with Hilbert space $\Hil_k$. These multiplicities can be calculated by
repeated use of the Clebsch-Gordan series for decomposing the tensor product of two
spin representations into irreducible components. In doing this for the chains $k$ and $k+1$,
in  each case starting by reducing the first two spins of the chain, then adding the third
and so on, one sees that each
spin-$S$ representation of the chain $k$, gives rise to a spin-$(S+\frac{1}{2})$ block
in the chain $k+1$. In case (I) this is so because the last step is adding a spin $\frac{1}{2}$.
In case (II) this is true because the last spin is of magnitude $\frac{1}{2}$ greater in the chain
$k+1$ than in the chain $k$. Hence, we have $d(k,S) \leq d(k+1,S+\frac{1}{2})$.
Moreover, it will be possible to choose the two bases
in such a way that
$$
(H_{k+1})_{ij}\leq (H_k)_{ij}\quad
\textrm{for all}\quad  1\leq i,j\leq d(k,S)\, .
$$
A Perron-Frobenius style theorem \cite{Wie,NSS} states that for any
two square matrices $A=(a_{ij})$, of size $m$, and $B=(b_{kl})$, of size $n$,
such that: (i) $n\geq m$, (ii) all off-diagonal matrix elements of $A$ and $B$
are non-postive, (iii) $b_{ij} \leq a_{ij}$ for all $1\leq i,j\leq m$, one
has $E(B)\leq E(A)$ where, as before, $E$ denotes the lowest eigenvalue.
The minimum eigenvalues are guaranteed to be simple if there exist $c\geq 0$
and an integer power $p\geq 1$ such that $(c{\bf I} - B)^p$ has all strictly
positive matrix elements. If in addition either $m<n$,
or at least one of the inequalities between the matrix elements is strict,
then $E(A)> E(B)$.
The inequality (\ref{mono}) is then obtained by applying this result with
$A=H_k$ and $B=H_{k+1}$.

Now, we construct the basis of the highest weight space.
One can think of the state space at site $x$ as the symmetric part of $2s_x$
spins-$\frac{1}{2}$. As the order of the spins $\frac{1}{2}$
is irrelevant, we can label the $2s_x+1$ states by the Ising configurations
\newcommand{\da}{\downarrow}
\newcommand{\ua}{\uparrow}
$$
\ket{\ua\ua\cdots\ua},\, \ket{\da\ua\cdots\ua},\,
\ket{\da\da\ua\cdots\ua},\, \ldots,\, \ket{\da\da\da\cdots\da}\, .
$$
E.g., $\ket{\downarrow\downarrow\uparrow\uparrow\uparrow}$
is the vector normally labelled as $\ket{j,m} = \ket{5/2,1/2}$, and {\em not}
the tensor $\ket{\da}\otimes\ket{\da}\otimes\ket{\ua}\otimes\ket{\ua}\otimes\ket{\ua}$.
The states for a chain of $L$ spins of magnitudes $s_1,\ldots,s_L$
are then tensor products of these configurations. We shall call such
vectors {\em ordered Ising configurations}.

\begin{figure}
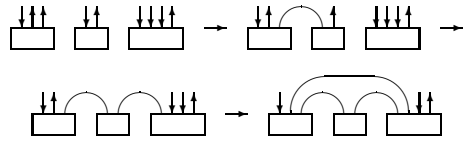

\resizebox{!}{2cm}{\egpic}
\caption{Construction of a basis vector from an ordered Ising
configuration}\label{fig:FKbasis}\end{figure}

Now, we construct a basis for the spin chain consisting
entirely of simultaneous eigenvectors of the total spin
and its third component, with eigenvalues $S$ and $M$, respectively.
Start from any ordered Ising configuration such that $2M=\#\ua -
\#\da$. Then, look for the leftmost $\da$   that has a $\ua$ to its left,
and draw an arc connecting this $\da$ to the rightmost $\ua$, left of it.
At this point, one may ignore the paired spins,
and repeat the procedure until there is no remaining unpaired $\da$ with
an unpaired $\ua$ to its left. This procedure guarantees that no arcs will
cross and no arc will span an unpaired spin. The result, when ignoring all paired
spins, is an ordered Ising  configuration of a single spin.
See Figure \ref{fig:FKbasis}
for an example of this procedure.
The total-spin $S$ is equal
to $\mathcal{S}$ minus the number of pairs.  Clearly, the highest
weight vectors are then those that have no unpaired $\da$.
The vectors
can be expanded in the tensor product basis by the following procedure:
each arc is replaced by the spin singlet $\ket{\ua}\otimes\ket{\da}-\ket{\da}
\otimes \ket{\ua}$, and the unpaired spins are replaced by their tensor
products. Finally, one symmetrizes in each block.
In the case $s_x=1/2$ for all $x$, this was called the Hulth\'en bracket
basis by Temperley and Lieb \cite{TL}, who also extended it to $\textrm{SU}_q(2)$
and proved that it is a basis. Frenkel and Khovanov introduced the general
case \cite{FK}.

Next, we calculate the matrix representation of the Hamiltonians
$H_k$ in the invariant subspaces of all highest weight vectors
of a given total spin. This is most easily accomplished by deriving a graphical
representation for the action of each term in the
Hamiltonian. We start with the case where all $s_x=1/2$. Then
the interaction terms generate the Temperley-Lieb algebra
with $q=1$ \cite{TL,KL}. The action of the negative of the interaction term in the
Hamiltonian can be graphically represented by
$$
-h_{x,x+1}=\frac{1}{2}(4 \boldsymbol{S}_x \cdot \boldsymbol{S}_{x+1}-1)
=  \begin{array}{c} \resizebox{1cm}{!}{\includegraphics{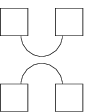}}\end{array} \, .
$$
We refer to the graphical diagram for $-h_{x,x+1}$ as $U_{x,x+1}$.
Let $\alpha$ be the diagram for the basis vector $\ket{\alpha}$.
The graphical rules are as follows: (i) if $x$ and $x+1$ are both unpaired
in $\alpha$, we have $h_{x,x+1}\ket{\alpha}=0$, (ii) if the composition of
$\alpha$ and $U_{x,x+1}$ is isotopic to $\beta$, with $\beta\neq\alpha$, then
the $h_{x,x+1}\ket{\alpha}=-\ket{\beta}$, (iii) if $\alpha=\beta$, then
$h_{x,x+1} \ket{\alpha} = 2 \ket{\alpha}$. This means that every loop contributes a factor
of $2$ to $h_{x,x+1}$. But this only happens when the ``cup'' of $U_{x,x+1}$ is
paired with an arc in $\alpha$. This is clearly a diagonal term. If the last site of an arc
for $\alpha$ is $x$, then there is no loop for $h_{y,y+1}$ with $y>x$.

To generalize this to arbitrary values of $s_x$, it is sufficient to
write the Heisenberg interaction for arbitrary spins as an interaction
between spin $\frac{1}{2}$ 's making up the spin $s_x$ and $s_{x+1}$, conjugated with
the projections onto the symmetric vectors.

The result is the following:
$$
-h_{x,x+1}
=\frac{1}{2}(\frac{1}{s_x s_{x+1}}\boldsymbol{S}_x \cdot \boldsymbol{S}_{x+1}-1)
= \begin{array}{c} \resizebox{1.85cm}{!}{\includegraphics{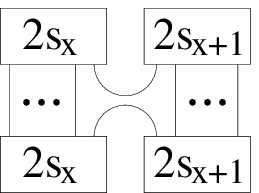}}
\end{array}\, .
$$
Here, the rectangles with label $2 s$ represent the symmetrizing projections
on the space of $2s$ spin $\frac{1}{2}$ variables.
The fundamental algebraic property
that will allow us to calculate the matrix elements of $H_k$ graphically
is the Jones-Wenzl relation (c.f., \cite{KL} and references therein):
\begin{equation}
\begin{array}{c} \resizebox{1cm}{!}{\includegraphics{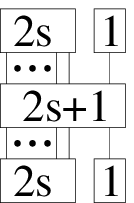}} \end{array}\, =\,
\begin{array}{c} \resizebox{1cm}{!}{\includegraphics{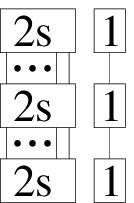}} \end{array}\, +\,
\frac{2s}{2s+1}\,
\begin{array}{c} \resizebox{1cm}{!}{\includegraphics{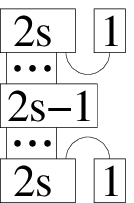}} \end{array}\, .
\end{equation}
For any element of the basis introduced above one can compute the action
of the Hamiltonian and write it as a linear combination of the same basis
vectors. From the grahical rules it is easy to observe that all off-diagonal matrix
elements are non-positive.

Furthermore, with each basis vector for the system $H_k$, we can identify
a basis vector for the system $H_{k+1}$. First, we consider the case where
one spin $1/2$ has been added to the system $H_k$. In that case one can simply
add one box with one up-spin to each basis vector. This then corresponds to
a basis vectors for the $H_{k+1}$ system where there is no arc to the last
spin. The case of increasing the magnitude of one of the spins by $1/2$
can be treated in a similar way by again using the Jones-Wenzl projection.
In this case, the label of the rightmost box in any basis vector for the
$H_k$ system is raised by one but the number of arcs remains unchanged.
Again, this clearly leads to a subset of the basis vectors for the
$H_{k+1}$ system.
The crucial property that allows us to compare the two Hamiltonians is the
following. When $H_{k+1}$ acts on a basis vector obtained from a corresponding
$H_k$ vector as we have just described, the only possible new
terms that are generated
are off-diagonal terms, which do not contain a bubble and, hence, are negative.
The details of the calculation of these matrix elements and further applications will appear
elsewhere \cite{NS}.

\section{Extensions and Discussion}\label{sec:discussion}

Our results can be extended in various directions. We discuss two
generalizations in particular.

The first is the analogous problem for the anisotropic model, i.e., the
ferromagnetic  XXZ model, with anisotropy $\Delta > 1$. Of course, the XXZ
Hamiltonian no longer has the full $\textrm{SU}(2)$ symmetry, but for the spin-$\frac{1}{2}$
chain there is an $\textrm{SU}_q(2)$-symmetric model obtained just by
adding suitable boundary fields \cite{PS}.
The representation theory of $\textrm{SU}_q(2)$ is
similar to that of $\textrm{SU}(2)$\cite{Kas}.
Using the $\textrm{SU}_q(2)$ symmetry, one can
show that, for a chain of $L$ sites, the first excited state has total
``spin'' $(L-1)/2$  \cite{KN}.
By replacing the Hulth\'en brackets with their $q-$analogues, FOEL
can be proved  for the spin-$\frac{1}{2}$ XXZ chains \cite{NSS}.
In principle one can generalize this result to higher spin models as well,
but one has to replace the standard XXZ interaction by a quantum group
symmetric interaction. Since such examples are of a lesser physical relevance we do not pursue this.

The second type of generalization concerns higher order interactions with the
full $\textrm{SU}(2)$ symmetry. As our argument to prove Theorem \ref{thm:main} only uses the symmetry of the Hamiltonian, and not its exact form, that result carries over immediately. The
dual canonical basis of highest weight vectors also does not depend on the Hamiltonian.
Theorem \ref{thm:mono} requires, however, that the off-diagonal elements of the Hamiltonian
with respect to this basis are non-negative. This puts a restriction on the
class of interactions. As an example, we consider the spin-1 chain.
For this case there is one non-trivial parameter
in the family of $\textrm{SU}(2)$-invariant nearest neighbor interactions, say the coefficient of the biquadratic term:
\begin{equation}
H=\sum_x (1-\boldsymbol{S}_x \cdot \boldsymbol{S}_{x+1}) + t(1-(\boldsymbol{S}_x \cdot \boldsymbol{S}_{x+1})^2)
\label{spin1}\end{equation}
It can be proved that the Hamiltonian (\ref{spin1}) has FOEL whenever $0\leq t\leq 1/3$,
and this result is optimal.

As a practical application of our results we point out the
following. Let $H$ be a Hamiltonian satisfying FOEL
and suppose one would like to find all the eigenvalues of $H$ below
energy $E$. It is sufficient to diagonalize $H$
in the subspaces with total spin $\mathcal{S} - k$, for $k=0,1,\ldots K$,
such that $E(H,\mathcal{S}-K) < E$.
The first value $K$ such that $E(H,\mathcal{S}-K-1) \geq E$ is large enough.
It is then guaranteed that one  has
{\em all} eigenvalues of the full $H$ below that value.
This is interesting because the dimension of the subspace for a chain of
length $L$ with total spin equal to $\mathcal{S}-k$ is
$O(L^{k})$, while the full Hilbert space has dimension $(2S+1)^L$
(when all particles have spin $S$).
In particular, one can obtain the gap by considering $K=1$.
%

\end{document}